\documentclass[useAMS,usenatbib]{mn2e}

\usepackage{threeparttable}

\usepackage{dpfloat}

\usepackage{amssymb}
\usepackage{mathptmx}       
\usepackage{helvet}         
\usepackage{courier}        
\usepackage{type1cm}        
\usepackage{graphicx}        
\usepackage{multicol}        
\usepackage[bottom]{footmisc}

\usepackage{natbib}
\usepackage{aas_macros}  

\usepackage{sidecap}
\usepackage{lscape}

\newcommand{\re}{$\mathrm{R_e}$}
\newcommand{\hb}{H$\beta$}
\newcommand{\hbo}{H$\beta_o$}
\newcommand{\sig}{$\sigma$}
\newcommand{\atlas}{ATLAS$\mathrm{^{3D}}$}

\newcommand{\rel}{[Mg/Fe] vs. [Fe/H]}

\title[The hELENa project -- II.]{The hELENa project -- II. Abundance distribution trends of early-type galaxies: from dwarfs to giants}

\author[Agnieszka Sybilska et al.]{A. Sybilska$^{1,2}$\thanks{E-mail: sybilska.aga@gmail.com}, H. Kuntschner$^{2}$, G. van de Ven$^{3}$, A. Vazdekis$^{4,5}$, J. Falc\'on-Barroso$^{4,5}$, \newauthor R.\,F. Peletier$^{6}$,  T. Lisker$^{7}$\\
$^{1}$Baltic Institute of Technology, al. Zwyci\c estwa 96/98, 81-451 Gdynia, Poland\\
$^{2}$European Southern Observatory, Karl-Schwarzschild-Strasse 2, 85748 Garching bei M{\"u}nchen, Germany\\
$^{3}$Max Planck Institute for Astronomy, K\"{o}nigstuhl 17, 69117 Heidelberg, Germany\\
$^{4}$Instituto de Astrof\'isica de Canarias, V\'ia L\'actea s/n, E-38205 La Laguna, Tenerife, Spain\\
$^{5}$Departamento de Astrof\'isica, Universidad de La Laguna, E-38200 La Laguna, Tenerife, Spain\\
$^{6}$Kapteyn Astronomical Institute, University of Groningen, Postbus 800, 9700 AV Groningen, the Netherlands\\
$^{7}$Astronomisches Rechen-Institut, Zentrum f\"{u}r Astronomie der Universit\"{a}t Heidelberg, M\"{o}nchhofstrasse 12-14, D-69120 Heidelberg, Germany\\
}

\begin{document}

\date{}


\maketitle

\label{firstpage}

\begin{abstract}

In this second paper of T\textbf{h}e role of \textbf{E}nvironment in shaping \textbf{L}ow-mass \textbf{E}arly-type \textbf{N}earby g\textbf{a}laxies (hELENa) series we study [Mg/Fe] abundance distribution trends of early-type galaxies observed with the SAURON integral field unit, spanning a wide range in mass and local environment densities: 20 low-mass early-types (dEs) of \cite{sybilska:2017} and 258 massive early types (ETGs) of the {\atlas} project, all homogeneously reduced and analyzed.  We show that the [Mg/Fe] ratios scale with velocity dispersion ({\sig}) at fixed [Fe/H] and that they evolve with [Fe/H] along similar paths for all early-types, grouped in bins of increasing local and global {\sig}, as well as the second velocity moment $V_{rms}$, indicating a common inside-out formation pattern. We then place our dEs on the {\rel} diagram of Local Group galaxies and show that dEs occupy the same region and show a similar trend line slope in the diagram as the high-metallicity stars of the Milky Way and the Large Magellanic Cloud. This finding extends the similar trend found for dwarf spheroidal $vs.$ dwarf irregular galaxies and supports the notion that dEs have evolved from late-type galaxies that have lost their gas at a point of their evolution, which likely coincided with them entering denser environments. 

\end{abstract}


\section{Introduction}

\begin{table*}
\caption{Basic details of the literature samples used in the study.}
 \begin{threeparttable}
\begin{centering}
\begin{tabular}{|r|r|r|r|r|}
\hline
data type & source & sample info & instrument & mass (range) $[M_{\odot}]$\\
\hline
unresolved/IFU & \protect\cite{sybilska:2017} & 20 dEs, Virgo and field & SAURON/WHT & 1.68$\cdot 10^{9}$ -- 8.18$\cdot 10^{9}$ $^{(a)}$ \\
 			& \protect\cite{scott:2013} & 258 ETGs, Virgo and field & SAURON/WHT & 3.86$\cdot 10^{9}$ -- 5.96$\cdot 10^{11}$ $^{(b)}$ \\

resolved & \protect\cite{tolstoy:2009}$^{(g)}$& Sculptor& UVES/VLT		 & 1.3$\pm$0.3, 2.3$\pm$0.2 $\cdot 10^{7}$ $^{(c)}$ \\
		 &							  		& Fornax    & UVES/VLT, FLAMES/VLT	 & 5.3$\pm$0.6, 7.4$\pm$0.4 $\cdot 10^{7}$ $^{(c)}$ \\
		 &							  		& Carina    & UVES/VLT		 & 0.6$\pm$0.2, 1.0$\pm$0.1 $\cdot 10^{7}$ $^{(c)}$ \\
		 & 							  		&Sagittarius& HIRES/Keck, FLAMES/VLT, UVES/VLT& 1 $\cdot 10^{9}$ $^{(d)}$ \\		 
		 & \protect\cite{kirby:2011a} 		& Sculptor  & DEIMOS/Keck 	 & 1.3$\pm$0.3, 2.3$\pm$0.2 $\cdot 10^{7}$ $^{(c)}$ \\
		 & 							 		& Leo I     & DEIMOS/Keck 	 & 1.2$\pm$0.3, 2.2$\pm$0.2 $\cdot 10^{7}$ $^{(c)}$ \\		 
		 & \protect\cite{pompeia:2008}		& LMC 	  & FLAMES/VLT 		 & 1.7$\pm$0.7 $\cdot 10^{10}$ $^{(e)}$\\
		 & \protect\cite{bensby:2014}   		& Milky Way & FEROS/ESO\,1.5m,\,2.2\,m; SOFIN,FIES/NOT;& 1.39$\pm$0.49 $\cdot 10^{12}$ $^{(f)}$ \\
		 & & & UVES/VLT; HARPS/ESO\,3.6\,m;  & \\
		 & & & MIKE/Magellan Clay telescope & \\
\hline		  				
\end{tabular}
\label{samples} 
\end{centering}
\end{threeparttable}

$^{(a)}$ $M_{vir}$ mass range based on {\re} and {\sig} values of \cite{sybilska:2017};
$^{(b)}$ $M_{JAM}$ mass range from \protect\cite{cappellari:2013a};
$^{(c)}$ $M_{1/2}$ from \cite{collins:2014} and \cite{wolf:2010}, respectively;
$^{(d)}$ Total mass of the best fitting cuspy/core model from \protect\cite{majewski:2013};
$^{(e)}$ M(8.7\,kpc) from \cite{marel:2014};
$^{(f)}$ M$_{100}$ mass of \cite{watkins:2010};
$^{(g)}$ compiled from \protect\cite{shetrone:2003}, \protect\cite{geisler:2005}, \protect\cite{mcwilliam:2005}, \protect\cite{letarte:2007}, \protect\cite{sborbone:2007}, and \protect\cite{koch:2008}.
\end{table*}

Detailed studies of galaxy properties as a function of redshift are hampered by the low signal-to-noise and low spatial resolution of high-redshift data. In order to study galaxy evolution we thus often turn to nearby galaxies to perform detailed studies of their stellar populations, with the hope to unveiling their evolutionary details through their star formation histories (SFHs). This so called ``stellar archaeology'' method has traditionally been limited by the age-metallicity degeneracy when deriving population properties from broadband colors. Using absorption line indices makes it possible, in principle, to lift the degeneracy, though here the challenge lies in ensuring that the derived metallicities are not affected by inherent abundance ratios. This, in turn, can be tackled by using abundance-ratio-independent index combinations (e.g. \citealt{kuntschner:2010}) and stellar population models specifically allowing for varying abundance ratios (e.g. \citealt{thomas:2005}).

The relation between element abundance ratios and metallicity has been studied in detail (through resolved observations) mostly for dwarf spheroidal galaxies (dSph) of the Milky Way (MW) due to their proximity to our Galaxy (see e.g. \citealt{kirby:2011b} or the review of \citealt{tolstoy:2009}). Some data also exist for Local Group (LG) dwarf irregulars, however, owing to their larger line-of-sight distances we are not able to study individual stars as faint as in the case of the closer dSphs, meaning the available samples do not cover comparably low metallicity regimes. 

Beyond the LG the situation is yet more complicated. Unable to obtain spectroscopic data for individual stars we need to resort to integrated measurements of stellar population parameters. These do not allow for the same level of detail to be examined since at each spatial point we get integrated, light-weighted information on all the underlying populations at that location. Nevertheless, with the help of stellar population models and various SFH recovery techniques we may still get insight into the assembly of those galaxies across the cosmic time. 

[Mg/Fe] abundance ratio is one of the fundamental measures of chemical enrichment in galaxies since different elements are produced in processes of different timescales, in this case type Ia versus type II supernovae (SNe), each having progenitors of different masses. Thus, the relation between [Mg/Fe] and [Fe/H] can provide information on the amount of feedback from the various SNe types and is expected to be constant for the lowest metallicity values, that is before type Ia SNe start contributing to the chemical evolution producing $\alpha$ elements. The downward turn of the {\rel} trend seen in the resolved MW data is known as a ``knee'' and is an indication of the metal enrichment achieved up to that point in time (i.e. how efficient SF was during the first $\sim$1\,Gyr of the galaxy evolution), thus, for example, a lower [Fe/H] at the knee location means lower feedback from star formation. This could be due to more mass being locked up in low-mass stars or, e.g. galactic winds causing outflow of metal-enhanced material (e.g. \citealt{tolstoy:2009}).

The knee location can only be determined for galaxies for which very metal-poor stars can be spectroscopically observed, which has so far only been achieved for the MW. Such metal-poor stars are still mostly out of reach for extragalactic objects, even those in the LG. Their large line-of-sight distances mean only the brightest stars can be observed, which dramatically limits the look-back time. Much outside of the LG we do not even have the luxury of individual star spectroscopic measurements and need to rely on integrated spectra. Theoretically, we would be able to obtain very low metallicity measurements only if the entire probed galaxy region was composed of the oldest, metal-poorest stars\footnote{We note, however, that with the use of techniques such as full-spectrum fitting (e.g. \protect\cite{koleva:2009b})one could, in principle, be able to detect an old, metal poor underlying population if it were sufficiently massive}.

Nevertheless, we can still benefit from the analysis of the {\rel} profiles and trends of various galaxy types across large mass ranges and different environments. Differences in the observed abundances contain clues as to the efficiency of internal enrichment mechanisms or the influence of environmental factors such as accretion or mergers. For example, a low [$\alpha$/Fe] at low [Fe/H] could indicate accretion of stellar material from objects where SF efficiency was lower (e.g. \citealt{wyse:2010}). The analysis of the loci of points and locations of the {\rel} profiles (at the high-metallicity end) can thus provide insight into differences between SFHs of the studied objects.

\section[Sample]{Sample selection}
\label{label:sample}
Below we provide a description of the literature samples used in the present study, which include integral-field unit (IFU) spectroscopic data for a total of 278 galaxies in the Virgo cluster and the group/field environment, as well as resolved data for MW and a number of LG galaxies. Individual references are provided below as well as in Tab.~\ref{samples}.\looseness-2

\subsection{Integrated-light data}

All our IFU data have been obtained with the SAURON IFU at the William Herschel Telescope in La Palma, Spain. The data reach out to ca. 1\,effective radius ({\re}) and  their spatial extent is limited either by field of view (FoV) of the SAURON instrument (for most of the ATLAS3D sample as well as some dEs) or the per-spaxel SNR (for dEs set to 7 and roughly corresponding to surface brightness of $\mu_V\approx23.5$\,mag at the edge of the field). For details on the sample selection and data reduction see the relevant papers cited below. Below we provide a short summary for the reader's convenience. \looseness-1

Our low-mass early-type sample consists of 20 dwarf early-types (\citealt{sybilska:2017}, hereafter S17), 17 of which are located in the Virgo cluster and three in the field. The galaxies have been drawn from the high-mass end of the dE luminosity function and comprise objects of various dE subtypes (disky, blue-core, as well as nucleated and non-nucleated galaxies), as well as a range of ellipticities and locations within the Virgo cluster.

The massive early-type sample consists of 258 {\atlas} galaxies: 58 Virgo cluster and 200 field/group objects. The population parameters presented here are, as described in S17, based on the line-strength measurements of \cite{scott:2013}, which were transformed from the Lick to the LIS system of \cite{vazdekis:2010} and for which stellar population estimates were then derived using MILES stellar population models (\citealt{vazdekis:2015}) in order to ensure maximal homogeneity. We also compare these population measurements to those for the same sample but derived with the use of \cite{schiavon:2007} models (Kuntschner et al., in prep.) to check for consistency -- see the appendix of S17 for details.

\subsection{Resolved data}

Data for LG dwarf spheroidal galaxies were taken from the works of \cite{tolstoy:2009} and \cite{kirby:2011b}. For these data sets we either directly took the provided moving averages/average profiles or calculated them ourselves. Additionally, we include the data from \cite{pompeia:2008} on the Large Magellanic Cloud (LMC) as well as  from \cite{bensby:2014} for the MW. We average the values of the latter so that an average trend line similar to those for early-types could be shown. The moving averages for the MW and the IFU sample were created by using a fixed subset with the size dependent on the number of available points.

\section{Methods}

\begin{figure*}
\centering
\includegraphics[width=2.13\columnwidth]{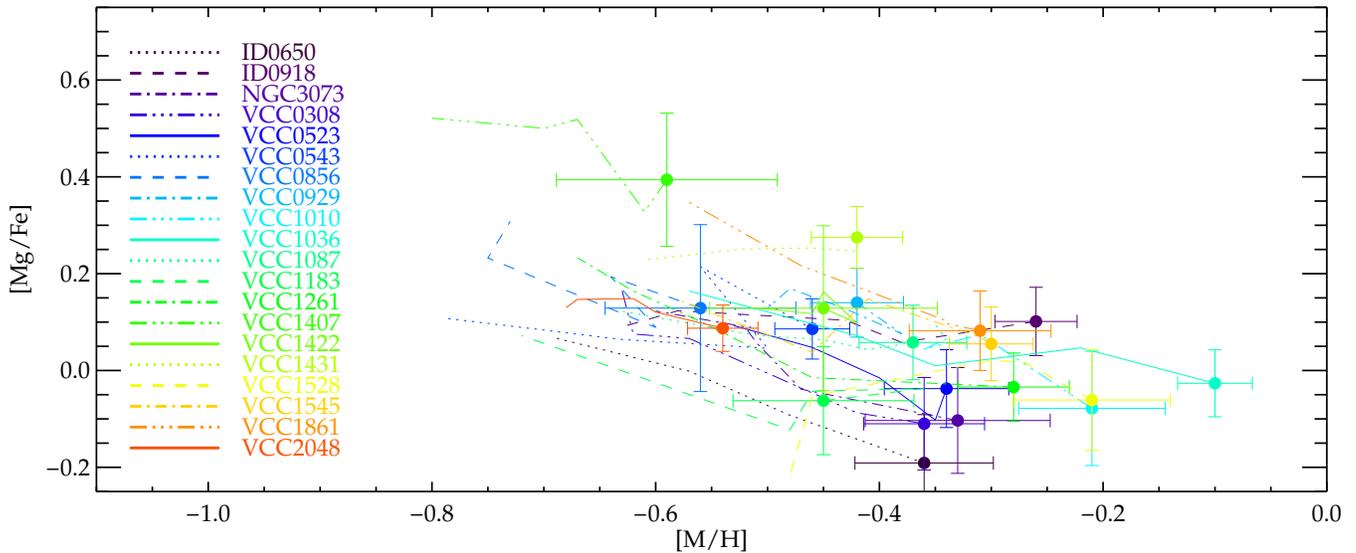}
\caption{[Mg/Fe] abundance ratio as a function of [M/H] of the 20 low-mass early-type galaxies of S17 from the Virgo cluster and the field/group environment. The galaxy centers have been marked with filled circles, error bars are shown for central points (the outermost profile points have error bars typically $\sim$twice as large as the central ones -- for the full profiles with uncertainties see the appendix of the above paper).}
\label{mgfe_vs_feh}  
\end{figure*}

\begin{figure*}
\centering
\includegraphics[width=2.13\columnwidth]{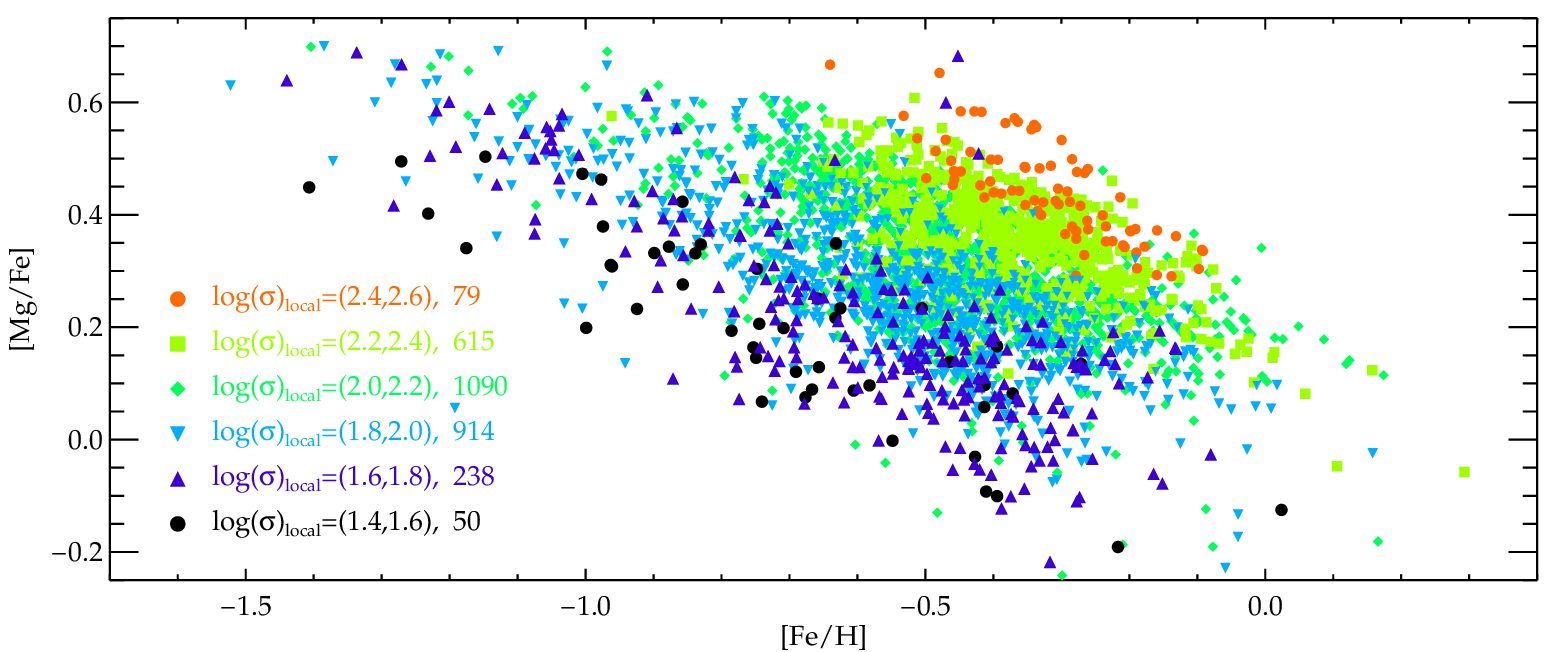}
\includegraphics[width=2.13\columnwidth]{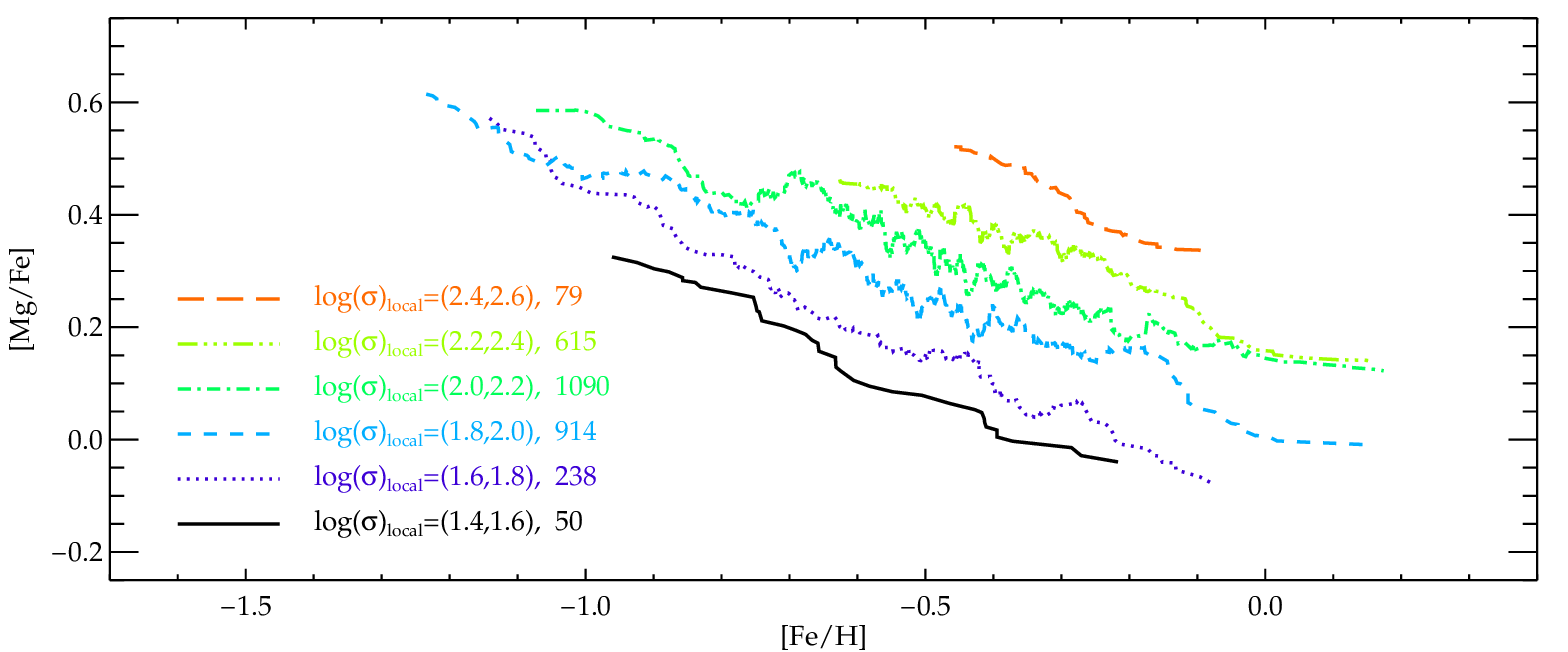}
\caption{\textbf{\underline{Upper panel:}} [Mg/Fe] abundance ratio as a function of [Fe/H] for all the profile points from the S17 sample as well as {\atlas} galaxies, color-coded by $\sigma_{local}$ values as explained in the legend, with the number of points falling into each bin provided next to the bin ranges. \textbf{\underline{Lower panel:}} running averages of all the profile points for the above defined {\sig} bins (same color-coding). [Fe/H] values have been obtained from [M/H] using the formula of \protect\cite{vazdekis:2015}: $[Fe/H] = [M/H] - 0.75 \cdot [Mg/Fe]$. Note that the running averages are \textit{not} averaged profiles as they combine points based on their [Fe/H] value and not location in a galaxy: for example, individual galaxies with flat [Mg/Fe] profiles are still able to produce a non-flat running average if their [Mg/Fe] and [Fe/H] values differ among the said galaxies; also, points belonging to one galaxy may be included in running averages for \textit{different} {\sig} bins.\looseness-2}
\label{mgfe_vs_feh2}  
\end{figure*}

\subsection{Deriving radial profiles}

Similar to stellar population profiles of S17, radial profiles of {\sig}, $V_{rms} = \sqrt{V^2+\sigma^2}$ were obtained from the S17 kinematic  maps by averaging bins\footnote{The maps were Voronoi binned to a min of S/N=30, see S17 for more details.} in elliptical annuli (i.e. along the lines of constant surface brightness) of increasing width, equal in log space. The same method was used by \citeauthor{kuntschner:2006} (\citeyear{kuntschner:2006}, \citeyear{kuntschner:2010}) for the SAURON survey and subsequently by \cite{scott:2013} for the ATLAS$^{3D}$ sample, with which we compare our results here. The errors on the averaged quantities were then obtained by taking a standard deviation of the values corresponding to the individual bins included in a given annulus. Population parameters were then derived from these annuli-averaged values.

\subsection{Determining stellar population parameters of the IFU samples}
The methods for the derivation of [M/H] and [Mg/Fe] abundance ratios are described in Sec.~3.6 and~3.7 of S17. In short, we use the abundance-ratio insensitive index combination [MgFe50]’ of \cite{kuntschner:2010} as well as the optimized {\hbo} index defined by \cite{cervantes:2009} which is less dependent on metallicity than the traditional {\hb} index. We then derive age and metallicity ([M/H]) with the help of MILES stellar population models of \cite{vazdekis:2015}, linearly interpolating between the available model prediction. To reduce the effects of grid discretization, we oversample the original models using a linear interpolation in the age-[M/H]-[index] space. [Fe/H] values are subsequently calculated using the following formula: $[Fe/H] = [M/H] – 0.75 \cdot [Mg/Fe]$ (ibid.).

To derive the [Mg/Fe] abundance ratios we interpolate between the predictions of MILES scaled-solar and $\alpha$-enhanced models on the Mg\,$b$-Fe5015 plane. For each line strength measurement we extracte Mg\,$b$ and Fe5015 index pairs from both sets of models, corresponding to the best-fitting single stellar population (SSP) age and metallicity derived by interpolating between model predictions in the {\hbo}-[MgFe50]' plane. We then calculate the distance between the two points as well as the distance between the points and our measured value to obtain the estimate of the [Mg/Fe] enhancement. \looseness-2 

While in principle the Fe5270 is more commonly chosen for the derivation of population parameters, the SAURON wavelength range is such that the index can only be measured for a small subsample of our galaxies (depending on their redshift). \citeauthor{kuntschner:2010} (\citeyear{kuntschner:2010}, their Fig.~4) have, however, shown for central apertures of their galaxies that the abundance ratios derived with the help of the two metal lines give consistent results.

We do note that the choice of models can influence the derived stellar population parameters values and introduce systematic bias. For example, a grid that is less orthogonal (i.e. shows a larger age-metallicity correlation) is able to systematically bias the derived parameters towards older ages or lower metallicities, and as a result influence the shape of the derived population profiles, an example of which can be seen in Fig. 3 of \cite{kuntschner:2010}. On the other hand, as shown in Fig. 6 of the same paper, metallicity and abundance ratio do not suffer from a degeneracy (error correlation) when analyzed together, hence the choice of these two stellar population parameters is justified and the {\rel} plane is an robust one to work with. This is further substantiated for the relative comparison of all our IFU data, given the same instrument and analysis approach for the entire sample.

For a comparison between the stellar population values obtained for the {\atlas} sample using the Lick system and \cite{schiavon:2007} models vs. the values presented here, please see the appendix of S17.

\subsection{Comparing resolved and unresolved observations}

\begin{figure*}
\centering
\includegraphics[width=2.13\columnwidth]{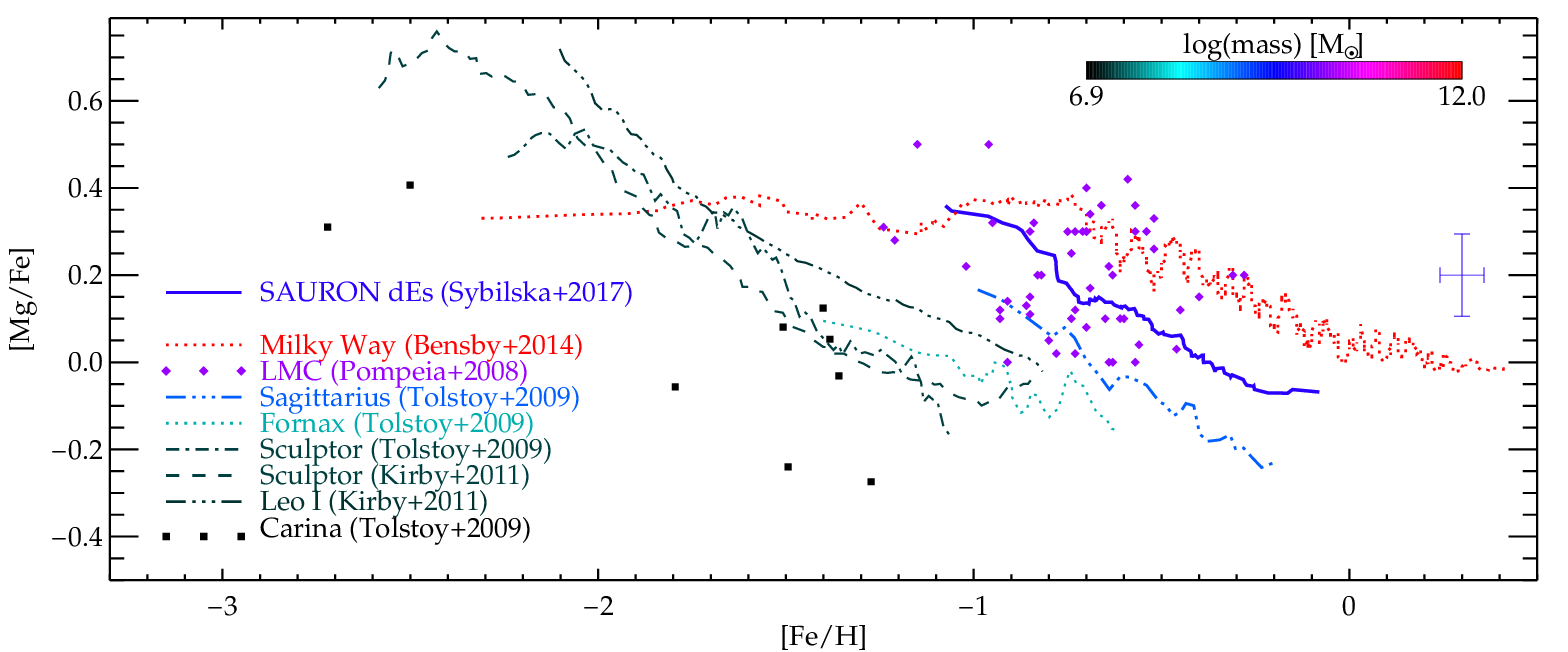}
\hspace{0.1cm}\includegraphics[width=2.08\columnwidth]{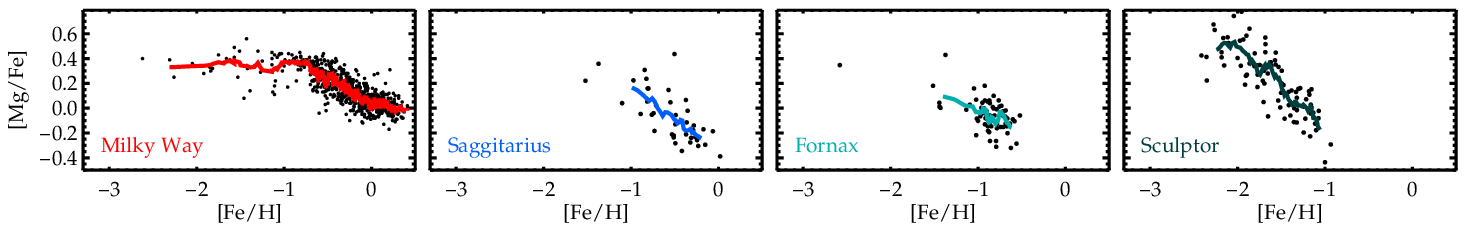}
\caption{\textbf{\underline{Upper panel:}}[Mg/Fe] abundance ratio as a function of [Fe/H] for the compilation of literature data on the Milky Way and Local Group dwarf galaxies, compared with our own data on low-mass early type galaxies . The galaxies are color-coded according to their dynamical mass as shown in the legend (see Tab.~\ref{label:sample} for details on the mass estimate sources and definitions) Blue cross shown on the right indicates a average error for the dE sample; for errors on individual galaxies see Fig. 1. \textbf{\underline{Lower panel:}} Individual measurements from \protect\cite{tolstoy:2009} and \protect\cite{bensby:2014}, overplotted with tracks shown in the upper panel.}
\label{mgfe_vs_feh-lit}  
\end{figure*}


\subsubsection{Globular cluster studies}
A number of recent studies have investigated the issue of comparability of metallicity and abundance ratio values obtained using integrated-light (IL) spectroscopic measurements versus high-resolution spectroscopy of individual stars in Galactic or Local Group globular clusters (GCs).

Based on a sample of 23 Galactic GCs \cite{pipino:2011} provide empirical calibrations for a conversion of Lick indices into abundances for the integrated light of old SSPs for a large range of observed [Fe/H] and [$\alpha$/H]. Their discussion on problems with the derivation of true [Mg/Fe] ratio applies only to the very low ($<$\,-1) [Fe/H] regime, not applicable in our case (except for 1 galaxy).   The expected underestimation of [Mg/Fe] is 0.05dex, which is within our measurement errors.

\cite{sakari:2016a} analyse 25 M31 GCs comparing results from their H-band integrated-light spectral measurements and high-resolution optical values from \cite{colucci:2014}. The relation showing the difference in the [Mg/Fe] values obtained using the two data sets as a function of [Fe/H] is shown in their Fig.\,8 and has a mean offset of -0.02 and a standard deviation of 0.16. Additionally, a comparison of their H-band results to low-resolution Lick index values from \cite{schiavon:2013} show a similarly good agreement between the two (mean offset: 0.0, standard deviation: 0.12).

We note that \cite{colucci:2017} find systematically lower (-0.24\,$\pm$\,0.07\,dex) [Mg/Fe] values from IL than from individual star measurements. However, the offset is less severe (ca. -0.15) for higher ($>$\,-1.0) [Fe/H] values. As a possible explanation of the discrepancy they give large abundance differences that might exist in a small fraction of luminous cool giants which dominate IL measurements but perhaps are not represented in the individual stars comparison sample.

The [Fe/H] values measured by \cite{larsen:2017} agree well with the literature values for individual stars from literature compilation (see their Table 4 for sources). For [Mg/Fe] they find that their IL values are systematically lower than those of \cite{carretta:2009} the average difference is ca. -0.15 dex, but again, the largest ($>$\,0.2\,dex) discrepancies are for the very low ($<$\,-2) values of [Fe/H], with the average offset for more metal-rich GCs being on average 0.1 dex. A comparison with other sources (\citealt{pritzl:2005}; \citealt{roediger:2014}) shows the two types of measurements to agree to within the errors. Additionally, the authors point out that even high-resolution spectroscopy of individual stars have systematic uncertainties of the order of ca. 0.1\,dex associated with them, hence an agreement of IL vs. individual stars measurements within this tolerance should be considered satisfactory.

\subsubsection{Galaxy studies}
More generally, the question of the feasibility of IL \textit{vs.} individual star measurements comparison has also been discussed in the context of dwarf as well as more massive galaxy studies. \cite{makarova:2010} compare resolved HST/ACS and integrated-light data from the 6\,m telescope of the Special Astrophysical Observatory of Russian Academy of Sciences for dSps/dEs in the M81 group and find that the two methods (CMD vs. full-spectrum fitting) give consistent [Fe/H] results for both galaxies ([Fe/H]\,$\approx$\,-\,1.5), albeit with some discrepancy for the younger, more metal-rich component in one of the objects). \looseness-2

In terms of much younger objects, \cite{garcia-benito-2012} compare SFHs for a nearby blue compact dwarf galaxy NGC\,6789 obtained through the analysis of HST/WFPC2 photometirc and WHT/ISIS spectroscopic data and find that they agree to within the errors. 

More recently, \cite{ruiz-lara:2015} carried out a similar test but expanded to systems with more complex star formation histories (SFHs). For a region in the LMC, they compare SFHs based on long-slit data from ESO 3.6\,m/EFOSC2 with SFHs based on photometry from HST/WFPC2. They conclude that a full-spectrum analysis of IL data is a reliable way of recovering SFHs even in SFH-wise complex systems, providing that a proper set of spectral templates is used. 

\section{Results}

\subsection{Virgo and field early-type galaxies}

Fig.~\ref{mgfe_vs_feh} shows the [Mg/Fe] abundance ratio as a function of [M/H] for each our dE galaxy individually. Looking at these profiles we see that the [Mg/Fe] ratios evolve with metallicity along nearly the same paths for all our dEs (see \citealt{kirby:2011b} for a similar result for LG dSphs). The values of [Mg/Fe] tend to increase outwards, though the associated gradients are mostly shallow or null to within the errors. [Fe/H] gradients, on the other hand, are all negative, with the majority \textit{not} negligible to within the errors. This has already been shown and quantified in S17 (Paper\,I of this series) and agrees with the literature (e.g. \citealt{koleva:2009b} who analyzed long-slit data for 16 dEs, belonging to the  Fornax cluster or nearby groups and find strong metallicity gradients for the majority of their sample, as well as \citealt{mentz:2016} who show MUSE integral-field unit data for one Virgo dE and find a gradient in both [Fe/H] and [Mg/Fe]).

Fig.~\ref{mgfe_vs_feh2} shows the {\rel} relation for all individual profile points (upper panel) as well as running averages (lower panel) for all early-type galaxies in our sample, grouped by local velocity dispersion $\sigma_{local}$. Note that, as also explained in the figure caption, these trend lines show averages for given [Fe/H] values, irrespective of the galaxy regions these correspond to (hence are \textit{not} profiles in the traditional sense), though as we move towards higher {\sig}, the same [Fe/H] \textit{do} typically correspond to regions further away from galaxy centers.

We see that the shape of these [Fe/H]-averaged trend lines is qualitatively similar for all early types, indicating comparable (inside-out) internal evolution. This has been shown earlier for the [Fe/H]-{\sig} and [Mg/Fe]-{\sig} relations separately for these samples (\citealt{mcdermid:2015}, \citealt{sybilska:2017}). However, even compared at fixed metallicity values, the [Mg/Fe] abundance ratios scale with velocity dispersion.  Here we also show that the massive galaxies' outskirts, while comparable in metallicity to the centers of dEs, are significantly more Mg-enhanced. On the other hand, similar levels of [Mg/Fe] are observed in, e.g. outskirts of low-{\sig} as in central parts of high-{\sig} galaxies, with the metallicities of the former significantly below those of the latter.  

\subsection{Virgo and field ETGs vs. Local Group late- and early-types}

Our findings expand on those of \cite{mentz:2016} who showed MUSE IFU data for one Virgo dE juxtaposed on literature data for the Local Group galaxies. Thanks to the much larger wavelength range available in MUSE, the authors were able to study not only Mg but also Na and Ca abundance ratios. In the present work, while limited to magnesium, we are able to show {\rel} profiles for a much larger sample of both low- and high-mass early types.\looseness-2

Fig.~\ref{mgfe_vs_feh-lit} shows the comparison between the dE galaxies of the previous subsection with the LG dwarfs as well as the MW. We have transformed the individual literature data points to the shown trends by calculating their running averages as described in Sec.~\ref{label:sample}.

We see that LG dEs/dSphs do not reach the [Fe/H] levels of more massive cluster dEs, which is expected from the mass-metallicity relation (see e.g. \citealt{tremonti:2004} or \citealt{gallazzi:2005} for relations derived from SDSS data for gas-phase and stellar metallicities, respectively -- which relations, however, do not reach the dSph low-mass levels; as well as \citealt{hidalgo:2017} for a $M_* - [Fe/H]_{RGB}$ relation for LG dwarfs covering the $10^3 - 10^{8.5} M_*$ mass range). 

To a first degree of approximation LG dSphs, do, however, show similar levels of [Mg/Fe] enhancement as more massive cluster dEs of the SAURON sample.  We also see that the shape of the {\rel} trend lines is similar for cluster dEs and LG dSphs which are up to 2 orders of magnitude less massive. This complements the findings of \cite{kirby:2011b} who found that MW dSph satellites' abundance ratios follow roughly the same path with increasing [Fe/H]. We note that our data do not reach as high [Mg/Fe] values as those of LG dSphs. This can possibly stem from the SNR limitations of the IFU dataset: if we were to probe regions much beyond 1\,{\re} (for which most of the integrated light would likely come from oldest i.e. high-[Mg/Fe] and low-[Fe/H] stars), it is possible that we would be seeing values of [Mg/Fe] as high as the above also for dEs.

Our dEs occupy roughly the same region of the {\rel} diagram as the LMC stars and the high-metallicity regions of the MW. A similar finding but for dSPh vs. dIrrs was shown in  \cite{tolstoy:2009} for  (see their Fig. 17), which they interpreted as dSphs being consistent with dIs that lost their gas at a late stage of their evolution.

\section{Discussion and Summary}

We have shown that the chemical enrichment histories of our galaxies resemble those of the MW disk\footnote{We note here that the thin and thick disks of the MW show different trends in [Fe/H]-[$\alpha$/Fe] diagrams (with [$\alpha$/Fe] being more enhanced in the latter), though the trend seems less pronounced in the [Mg/Fe]-[Fe/H] relation itself (see, e.g. \protect\citealt{bensby:2014}, sec. 6, Fig. 15 and \protect\citealt{bland-hawthorn:2016}, sec. 5.2.2., and references therein.} and the LG late-type dwarf LMC in that they occupy the same region in the {\rel} diagram and the slopes of their {\rel} trend lines are similar. Following up on the \cite{tolstoy:2009} finding for dSphs and dIrrs discussed earlier, we could argue that dEs extend the trend for late-type galaxies of comparable or higher masses and thus support the notion that dEs have evolved from LTGs which have lost their star-forming material at a point in their evolutionary history, which likely coincided with entering regions of higher environmental density (either groups or clusters). However, with the current data we are unable to put a time stamp on such an event happening.

Low-mass ETGs are known to have lower [Mg/Fe] ratios than massive ETGs (\citealt{michielsen:2008}, S17) but an even stronger relation exists between mass/$\sigma$ and metallicity. Thanks to our spatially resolved data, i.e. the availability of local [Fe/H] and [Mg/Fe] values, we are able to show that the relation between $\sigma$ and [Mg/Fe] holds even at fixed [Fe/H]. Lower [Mg/Fe] at a given [Fe/H] in the case of dEs and LMC indicates an overall less efficient enrichment mechanism(s) as compared with, e.g. massive ETGs, since such galaxies have not managed to produce large amounts of metal relative to the available star-building material. This means lower-mass galaxies (exhibiting low [Mg/Fe] ratios) are either inefficient at turning gas into stars or they do produce metals but are unable to retain them (and so the new generations of stars are not enriched), either due to galactic winds, or environmental factors such as tidal forces or starvation which are able to stop the star forming process. 

For galaxies of similar masses, differences in [Mg/Fe] ratios at fixed [Fe/H] can provide clues on the galaxies’ merger histories. Zolotov et al. (2010) find in their simulations that accreted stars, coming from lower-mass objects, have lower [O/Fe] values (used as proxy for [alpha//Fe]) at the high-[Fe/H] end than those stars that were formed in situ. The authors show that differences in [O/Fe] at similar [Fe/H] result from the different potential wells within which in situ and accreted halo stars formed. While the various [$\alpha$//Fe] vs. [Fe/H] relations might differ quantitatively, they all show the same (qualitative) behaviour in that the shape of the above relation shows a downturn at high [Fe/H] ratios (e.g. \citealt{tolstoy:2009}, \citealt{mentz:2016}, etc.) Therefore, for galaxies of equal masses to have similar [$\alpha$/Fe] ratios at fixed [Fe/H] would mean that they likely have had similar merger histories - their stars having formed in similarly deep potential wells (unless the processes that affected the ratios managed to cancel each other out): alternatively, a galaxy from the above hypothetical pair but with lower [Mg/Fe], i.e. with. a larger portion of its mass contained in stars of lower [Mg/Fe] must have obtained them from smaller-mass galaxies through a (series of) minor merger(s). We do not find a significant difference between dEs and LMC, suggesting that -- in terms of merger histories -- their evolutionary paths could have been similar. 

The results presented here are in agreement with those of \cite{mentz:2016} for one dE observed with MUSE, as well as those in \cite{sen:2017} who show central [Mg/Fe] values for 11 Virgo dEs drawn from the SMAKCED project sample \citep{toloba:2014b}. Besides [Mg/Fe], these works have been able to examine [Ca/Fe] and [Na/Fe] abundance ratios and noted the difference of the former between their dEs and the LMC. However, a detailed discussion of discrepancies seen in different elements' abundance distributions is beyond the scope of this paper.\looseness-2

Our approach presented here is limited by the SAURON wavelength range to the Fe5015 and Mg\,$b$ metallicity-tracing indices. To strengthen and expand on our findings, as well as those quoted above, it would therefore be advisable to obtain for our sample spectroscopic data covering a larger wavelength range (such as that available with MUSE) so that other $\alpha$ elements can be studied simultaneously alongside magnesium. 

\section*{Acknowledgments}

We thank the anonymous referee whose suggestions and comments helped improve the proesentation of the paper results. AS acknowledges the suuport of the Alexander von Humboldt Foundation through a Humboldt Research Fellowship for Postdoctoral Researchers. The paper is based on observations obtained at the William Herschel Telescope, operated by the Isaac Newton Group in the Spanish Observatorio del Roque de los Muchachos of the Instituto de Astrof{\'i}sica de Canarias. We thank Nicholas Scott for making his local velocity dipersion measurements for the {\atlas} sample available to us. AV and JFB acknowledge support from grant AYA2016-77237-C3-1-P from the Spanish Ministry of Economy and Competitiveness (MINECO). RFP and TL acknowledge financial support from the European Union's Horizon 2020 research and innovation programme under the Marie Sk{\l}odowska-Curie grant agreement No 721463 to the SUNDIAL ITN network.

\bibliography{biblio}

\begin{thebibliography}{}

\bibitem[\protect\citeauthoryear{{Bensby}, {Feltzing} \& {Oey}}{{Bensby}
  et~al.}{2014}]{bensby:2014}
{Bensby} T.,  {Feltzing} S.,    {Oey} M.~S.,  2014, \aap, 562, A71

\bibitem[\protect\citeauthoryear{{Bland-Hawthorn} \&
  {Gerhard}}{{Bland-Hawthorn} \& {Gerhard}}{2016}]{bland-hawthorn:2016}
{Bland-Hawthorn} J.,  {Gerhard} O.,  2016, \araa, 54, 529

\bibitem[\protect\citeauthoryear{{{\c S}en}, {Peletier}, {Boselli}, {den Brok},
  {Falc{\'o}n-Barroso}, {Hensler}, {Janz}, {Laurikainen}, {Lisker}, {Mentz},
  {Paudel}, {Salo} \& {Sybilska}}{{{\c S}en} et~al.}{2017}]{sen:2017}
{{\c S}en} {\c S}.,  {Peletier} R.~F.,  {Boselli} A.,  {den Brok} M.,
  {Falc{\'o}n-Barroso} J.,  {Hensler} G.,  {Janz} J.,  {Laurikainen} E.,
  {Lisker} T.,  {Mentz} J.~J.,  {Paudel} S.,  {Salo} H.,    {Sybilska} A.,
  2017, ArXiv e-prints

\bibitem[\protect\citeauthoryear{{Cappellari}, {Scott}, {Alatalo}, {Blitz},
  {Bois}, {Bournaud}, {Bureau}, {Crocker}, {Davies}, {Davis}, {de Zeeuw},
  {Duc}, {Emsellem}, {Khochfar}, {Krajnovi{\'c}}, {Kuntschner} \&
  {McDermid}}{{Cappellari} et~al.}{2013}]{cappellari:2013a}
{Cappellari} M.,  {Scott} N.,  {Alatalo} K.,  {Blitz} L.,  {Bois} M.,
  {Bournaud} F.,  {Bureau} M.,  {Crocker} A.~F.,  {Davies} R.~L.,  {Davis}
  T.~A.,  {de Zeeuw} P.~T.,  {Duc} P.-A.,  {Emsellem} E.,  {Khochfar} S.,
  {Krajnovi{\'c}} D.,  {Kuntschner} H.,    {McDermid} R.~M.,  2013, \mnras

\bibitem[\protect\citeauthoryear{{Carretta}, {Bragaglia}, {Gratton} \&
  {Lucatello}}{{Carretta} et~al.}{2009}]{carretta:2009}
{Carretta} E.,  {Bragaglia} A.,  {Gratton} R.,    {Lucatello} S.,  2009, \aap,
  505, 139

\bibitem[\protect\citeauthoryear{{Cervantes} \& {Vazdekis}}{{Cervantes} \&
  {Vazdekis}}{2009}]{cervantes:2009}
{Cervantes} J.~L.,  {Vazdekis} A.,  2009, \mnras, 392, 691

\bibitem[\protect\citeauthoryear{{Collins}, {Chapman}, {Rich}, {Ibata},
  {Martin}, {Irwin}, {Bate}, {Lewis}, {Pe{\~n}arrubia}, {Arimoto}, {Casey},
  {Ferguson}, {Koch}, {McConnachie} \& {Tanvir}}{{Collins}
  et~al.}{2014}]{collins:2014}
{Collins} M.~L.~M.,  {Chapman} S.~C.,  {Rich} R.~M.,  {Ibata} R.~A.,  {Martin}
  N.~F.,  {Irwin} M.~J.,  {Bate} N.~F.,  {Lewis} G.~F.,  {Pe{\~n}arrubia} J.,
  {Arimoto} N.,  {Casey} C.~M.,  {Ferguson} A.~M.~N.,  {Koch} A.,
  {McConnachie} A.~W.,    {Tanvir} N.,  2014, \apj, 783, 7

\bibitem[\protect\citeauthoryear{{Colucci}, {Bernstein} \& {Cohen}}{{Colucci}
  et~al.}{2014}]{colucci:2014}
{Colucci} J.~E.,  {Bernstein} R.~A.,    {Cohen} J.~G.,  2014, \apj, 797, 116

\bibitem[\protect\citeauthoryear{{Colucci}, {Bernstein} \&
  {McWilliam}}{{Colucci} et~al.}{2017}]{colucci:2017}
{Colucci} J.~E.,  {Bernstein} R.~A.,    {McWilliam} A.,  2017, \apj, 834, 105

\bibitem[\protect\citeauthoryear{{Gallazzi}, {Charlot}, {Brinchmann}, {White}
  \& {Tremonti}}{{Gallazzi} et~al.}{2005}]{gallazzi:2005}
{Gallazzi} A.,  {Charlot} S.,  {Brinchmann} J.,  {White} S.~D.~M.,
  {Tremonti} C.~A.,  2005, \mnras, 362, 41

\bibitem[\protect\citeauthoryear{{Garc{\'{\i}}a-Benito} \&
  {P{\'e}rez-Montero}}{{Garc{\'{\i}}a-Benito} \&
  {P{\'e}rez-Montero}}{2012}]{garcia-benito-2012}
{Garc{\'{\i}}a-Benito} R.,  {P{\'e}rez-Montero} E.,  2012, \mnras, 423, 406

\bibitem[\protect\citeauthoryear{{Geisler}, {Smith}, {Wallerstein}, {Gonzalez}
  \& {Charbonnel}}{{Geisler} et~al.}{2005}]{geisler:2005}
{Geisler} D.,  {Smith} V.~V.,  {Wallerstein} G.,  {Gonzalez} G.,
  {Charbonnel} C.,  2005, \aj, 129, 1428

\bibitem[\protect\citeauthoryear{{Hidalgo}}{{Hidalgo}}{2017}]{hidalgo:2017}
{Hidalgo} S.~L.,  2017, ArXiv e-prints

\bibitem[\protect\citeauthoryear{{Kirby}, {Cohen}, {Smith}, {Majewski}, {Sohn}
  \& {Guhathakurta}}{{Kirby} et~al.}{2011}]{kirby:2011b}
{Kirby} E.~N.,  {Cohen} J.~G.,  {Smith} G.~H.,  {Majewski} S.~R.,  {Sohn}
  S.~T.,    {Guhathakurta} P.,  2011, \apj, 727, 79

\bibitem[\protect\citeauthoryear{{Kirby}, {Lanfranchi}, {Simon}, {Cohen} \&
  {Guhathakurta}}{{Kirby} et~al.}{2011}]{kirby:2011a}
{Kirby} E.~N.,  {Lanfranchi} G.~A.,  {Simon} J.~D.,  {Cohen} J.~G.,
  {Guhathakurta} P.,  2011, \apj, 727, 78

\bibitem[\protect\citeauthoryear{{Koch}, {Grebel}, {Gilmore}, {Wyse}, {Kleyna},
  {Harbeck}, {Wilkinson} \& {Wyn Evans}}{{Koch} et~al.}{2008}]{koch:2008}
{Koch} A.,  {Grebel} E.~K.,  {Gilmore} G.~F.,  {Wyse} R.~F.~G.,  {Kleyna}
  J.~T.,  {Harbeck} D.~R.,  {Wilkinson} M.~I.,    {Wyn Evans} N.,  2008, \aj,
  135, 1580

\bibitem[\protect\citeauthoryear{{Koleva}, {de Rijcke}, {Prugniel}, {Zeilinger}
  \& {Michielsen}}{{Koleva} et~al.}{2009}]{koleva:2009b}
{Koleva} M.,  {de Rijcke} S.,  {Prugniel} P.,  {Zeilinger} W.~W.,
  {Michielsen} D.,  2009, \mnras, 396, 2133

\bibitem[\protect\citeauthoryear{{Kuntschner}, {Emsellem}, {Bacon}, {Bureau},
  {Cappellari}, {Davies}, {de Zeeuw}, {Falc{\'o}n-Barroso}, {Krajnovi{\'c}},
  {McDermid}, {Peletier} \& {Sarzi}}{{Kuntschner}
  et~al.}{2006}]{kuntschner:2006}
{Kuntschner} H.,  {Emsellem} E.,  {Bacon} R.,  {Bureau} M.,  {Cappellari} M.,
  {Davies} R.~L.,  {de Zeeuw} P.~T.,  {Falc{\'o}n-Barroso} J.,  {Krajnovi{\'c}}
  D.,  {McDermid} R.~M.,  {Peletier} R.~F.,    {Sarzi} M.,  2006, \mnras, 369,
  497

\bibitem[\protect\citeauthoryear{{Kuntschner}, {Emsellem}, {Bacon},
  {Cappellari}, {Davies}, {de Zeeuw}, {Falc{\'o}n-Barroso}, {Krajnovi{\'c}},
  {McDermid}, {Peletier}, {Sarzi}, {Shapiro}, {van den Bosch} \& {van de
  Ven}}{{Kuntschner} et~al.}{2010}]{kuntschner:2010}
{Kuntschner} H.,  {Emsellem} E.,  {Bacon} R.,  {Cappellari} M.,  {Davies}
  R.~L.,  {de Zeeuw} P.~T.,  {Falc{\'o}n-Barroso} J.,  {Krajnovi{\'c}} D.,
  {McDermid} R.~M.,  {Peletier} R.~F.,  {Sarzi} M.,  {Shapiro} K.~L.,  {van den
  Bosch} R.~C.~E.,    {van de Ven} G.,  2010, \mnras, 408, 97

\bibitem[\protect\citeauthoryear{{Larsen}, {Brodie} \& {Strader}}{{Larsen}
  et~al.}{2017}]{larsen:2017}
{Larsen} S.~S.,  {Brodie} J.~P.,    {Strader} J.,  2017, \aap, 601, A96

\bibitem[\protect\citeauthoryear{{Letarte}}{{Letarte}}{2007}]{letarte:2007}
{Letarte} B.,  2007, PhD thesis, University of Groningen

\bibitem[\protect\citeauthoryear{{Majewski}, {Hasselquist}, {{\L}okas},
  {Nidever}, {Frinchaboy}, {Garc{\'{\i}}a P{\'e}rez}, {Johnston} \&
  {M{\'e}sz{\'a}ros}}{{Majewski} et~al.}{2013}]{majewski:2013}
{Majewski} S.~R.,  {Hasselquist} S.,  {{\L}okas} E.~L.,  {Nidever} D.~L.,
  {Frinchaboy} P.~M.,  {Garc{\'{\i}}a P{\'e}rez} A.~E.,  {Johnston} K.~V.,
  {M{\'e}sz{\'a}ros} S.,  2013, \apjl, 777, L13

\bibitem[\protect\citeauthoryear{{Makarova}, {Koleva}, {Makarov} \&
  {Prugniel}}{{Makarova} et~al.}{2010}]{makarova:2010}
{Makarova} L.,  {Koleva} M.,  {Makarov} D.,    {Prugniel} P.,  2010, \mnras,
  406, 1152

\bibitem[\protect\citeauthoryear{{McDermid}, {Alatalo}, {Blitz}, {Bournaud},
  {Bureau}, {Cappellari}, {Crocker}, {Davies}, {Davis}, {de Zeeuw}, {Duc},
  {Emsellem}, {Khochfar} \& {Krajnovi{\'c}}}{{McDermid}
  et~al.}{2015}]{mcdermid:2015}
{McDermid} R.~M.,  {Alatalo} K.,  {Blitz} L.,  {Bournaud} F.,  {Bureau} M.,
  {Cappellari} M.,  {Crocker} A.~F.,  {Davies} R.~L.,  {Davis} T.~A.,  {de
  Zeeuw} P.~T.,  {Duc} P.-A.,  {Emsellem} E.,  {Khochfar} S.,
  {Krajnovi{\'c}} D.,  2015, \mnras, 448, 3484

\bibitem[\protect\citeauthoryear{{McWilliam} \& {Smecker-Hane}}{{McWilliam} \&
  {Smecker-Hane}}{2005}]{mcwilliam:2005}
{McWilliam} A.,  {Smecker-Hane} T.~A.,  2005, \apjl, 622, L29

\bibitem[\protect\citeauthoryear{{Mentz}, {La Barbera}, {Peletier},
  {Falc{\'o}n-Barroso}, {Lisker}, {van de Ven}, {Loubser} \& {Hilker}}{{Mentz}
  et~al.}{2016}]{mentz:2016}
{Mentz} J.~J.,  {La Barbera} F.,  {Peletier} R.~F.,  {Falc{\'o}n-Barroso} J.,
  {Lisker} T.,  {van de Ven} G.,  {Loubser} S.~I.,    {Hilker} M.,  2016,
  \mnras, 463, 2819

\bibitem[\protect\citeauthoryear{{Michielsen}, {Boselli}, {Conselice},
  {Toloba}, {Whiley}, {Arag{\'o}n-Salamanca}, {Balcells}, {Cardiel}, {Cenarro},
  {Gorgas}, {Peletier} \& {Vazdekis}}{{Michielsen}
  et~al.}{2008}]{michielsen:2008}
{Michielsen} D.,  {Boselli} A.,  {Conselice} C.~J.,  {Toloba} E.,  {Whiley}
  I.~M.,  {Arag{\'o}n-Salamanca} A.,  {Balcells} M.,  {Cardiel} N.,  {Cenarro}
  A.~J.,  {Gorgas} J.,  {Peletier} R.~F.,    {Vazdekis} A.,  2008, \mnras, 385,
  1374

\bibitem[\protect\citeauthoryear{{Pipino} \& {Danziger}}{{Pipino} \&
  {Danziger}}{2011}]{pipino:2011}
{Pipino} A.,  {Danziger} I.~J.,  2011, \aap, 530, A22

\bibitem[\protect\citeauthoryear{{Pomp{\'e}ia}, {Hill}, {Spite}, {Cole},
  {Primas}, {Romaniello}, {Pasquini}, {Cioni} \& {Smecker Hane}}{{Pomp{\'e}ia}
  et~al.}{2008}]{pompeia:2008}
{Pomp{\'e}ia} L.,  {Hill} V.,  {Spite} M.,  {Cole} A.,  {Primas} F.,
  {Romaniello} M.,  {Pasquini} L.,  {Cioni} M.-R.,    {Smecker Hane} T.,  2008,
  \aap, 480, 379

\bibitem[\protect\citeauthoryear{{Pritzl}, {Venn} \& {Irwin}}{{Pritzl}
  et~al.}{2005}]{pritzl:2005}
{Pritzl} B.~J.,  {Venn} K.~A.,    {Irwin} M.,  2005, \aj, 130, 2140

\bibitem[\protect\citeauthoryear{{Roediger}, {Courteau}, {Graves} \&
  {Schiavon}}{{Roediger} et~al.}{2014}]{roediger:2014}
{Roediger} J.~C.,  {Courteau} S.,  {Graves} G.,    {Schiavon} R.~P.,  2014,
  \apjs, 210, 10

\bibitem[\protect\citeauthoryear{{Ruiz-Lara}, {P{\'e}rez}, {Gallart}, {Alloin},
  {Monelli}, {Koleva}, {Pompei}, {Beasley}, {S{\'a}nchez-Bl{\'a}zquez},
  {Florido}, {Aparicio}, {Fleurence}, {Hardy}, {Hidalgo} \&
  {Raimann}}{{Ruiz-Lara} et~al.}{2015}]{ruiz-lara:2015}
{Ruiz-Lara} T.,  {P{\'e}rez} I.,  {Gallart} C.,  {Alloin} D.,  {Monelli} M.,
  {Koleva} M.,  {Pompei} E.,  {Beasley} M.,  {S{\'a}nchez-Bl{\'a}zquez} P.,
  {Florido} E.,  {Aparicio} A.,  {Fleurence} E.,  {Hardy} E.,  {Hidalgo} S.,
  {Raimann} D.,  2015, \aap, 583, A60

\bibitem[\protect\citeauthoryear{{Sakari} \& {Wallerstein}}{{Sakari} \&
  {Wallerstein}}{2016}]{sakari:2016a}
{Sakari} C.~M.,  {Wallerstein} G.,  2016, \mnras, 456, 831

\bibitem[\protect\citeauthoryear{{Sbordone}, {Bonifacio}, {Buonanno},
  {Marconi}, {Monaco} \& {Zaggia}}{{Sbordone} et~al.}{2007}]{sborbone:2007}
{Sbordone} L.,  {Bonifacio} P.,  {Buonanno} R.,  {Marconi} G.,  {Monaco} L.,
  {Zaggia} S.,  2007, \aap, 465, 815

\bibitem[\protect\citeauthoryear{{Schiavon}}{{Schiavon}}{2007}]{schiavon:2007}
{Schiavon} R.~P.,  2007, \apjs, 171, 146

\bibitem[\protect\citeauthoryear{{Schiavon}, {Caldwell}, {Conroy}, {Graves},
  {Strader}, {MacArthur}, {Courteau} \& {Harding}}{{Schiavon}
  et~al.}{2013}]{schiavon:2013}
{Schiavon} R.~P.,  {Caldwell} N.,  {Conroy} C.,  {Graves} G.~J.,  {Strader} J.,
   {MacArthur} L.~A.,  {Courteau} S.,    {Harding} P.,  2013, \apjl, 776, L7

\bibitem[\protect\citeauthoryear{{Scott}, {Cappellari}, {Davies}, {Kleijn},
  {Bois}, {Alatalo}, {Blitz}, {Bournaud}, {Bureau}, {Crocker}, {Davis}, {de
  Zeeuw}, {Duc} \& {Emsellem}}{{Scott} et~al.}{2013}]{scott:2013}
{Scott} N.,  {Cappellari} M.,  {Davies} R.~L.,  {Kleijn} G.~V.,  {Bois} M.,
  {Alatalo} K.,  {Blitz} L.,  {Bournaud} F.,  {Bureau} M.,  {Crocker} A.,
  {Davis} T.~A.,  {de Zeeuw} P.~T.,  {Duc} P.-A.,    {Emsellem} E.,  2013,
  \mnras, 432, 1894

\bibitem[\protect\citeauthoryear{{Shetrone}, {Venn}, {Tolstoy}, {Primas},
  {Hill} \& {Kaufer}}{{Shetrone} et~al.}{2003}]{shetrone:2003}
{Shetrone} M.,  {Venn} K.~A.,  {Tolstoy} E.,  {Primas} F.,  {Hill} V.,
  {Kaufer} A.,  2003, \aj, 125, 684

\bibitem[\protect\citeauthoryear{{Sybilska}, {Lisker}, {Kuntschner},
  {Vazdekis}, {van de Ven}, {Peletier}, {Falc{\'o}n-Barroso}, {Vijayaraghavan}
  \& {Janz}}{{Sybilska} et~al.}{2017}]{sybilska:2017}
{Sybilska} A.,  {Lisker} T.,  {Kuntschner} H.,  {Vazdekis} A.,  {van de Ven}
  G.,  {Peletier} R.,  {Falc{\'o}n-Barroso} J.,  {Vijayaraghavan} R.,    {Janz}
  J.,  2017, \mnras, 470, 815

\bibitem[\protect\citeauthoryear{{Thomas}, {Maraston}, {Bender} \& {Mendes de
  Oliveira}}{{Thomas} et~al.}{2005}]{thomas:2005}
{Thomas} D.,  {Maraston} C.,  {Bender} R.,    {Mendes de Oliveira} C.,  2005,
  \apj, 621, 673

\bibitem[\protect\citeauthoryear{{Toloba}, {Guhathakurta}, {Peletier},
  {Boselli}, {Lisker}, {Falc{\'o}n-Barroso}, {Simon} \& {van de Ven}}{{Toloba}
  et~al.}{2014}]{toloba:2014b}
{Toloba} E.,  {Guhathakurta} P.,  {Peletier} R.~F.,  {Boselli} A.,  {Lisker}
  T.,  {Falc{\'o}n-Barroso} J.,  {Simon} J.~D.,    {van de Ven} G.,  2014,
  \apjs, 215, 17

\bibitem[\protect\citeauthoryear{{Tolstoy}, {Hill} \& {Tosi}}{{Tolstoy}
  et~al.}{2009}]{tolstoy:2009}
{Tolstoy} E.,  {Hill} V.,    {Tosi} M.,  2009, \araa, 47, 371

\bibitem[\protect\citeauthoryear{{Tremonti}, {Heckman}, {Kauffmann},
  {Brinchmann}, {Charlot}, {White}, {Seibert}, {Peng}, {Schlegel}, {Uomoto},
  {Fukugita} \& {Brinkmann}}{{Tremonti} et~al.}{2004}]{tremonti:2004}
{Tremonti} C.~A.,  {Heckman} T.~M.,  {Kauffmann} G.,  {Brinchmann} J.,
  {Charlot} S.,  {White} S.~D.~M.,  {Seibert} M.,  {Peng} E.~W.,  {Schlegel}
  D.~J.,  {Uomoto} A.,  {Fukugita} M.,    {Brinkmann} J.,  2004, \apj, 613, 898

\bibitem[\protect\citeauthoryear{{van der Marel} \& {Kallivayalil}}{{van der
  Marel} \& {Kallivayalil}}{2014}]{marel:2014}
{van der Marel} R.~P.,  {Kallivayalil} N.,  2014, \apj, 781, 121

\bibitem[\protect\citeauthoryear{{Vazdekis}, {Coelho}, {Cassisi},
  {Ricciardelli}, {Falc{\'o}n-Barroso}, {S{\'a}nchez-Bl{\'a}zquez}, {Barbera},
  {Beasley} \& {Pietrinferni}}{{Vazdekis} et~al.}{2015}]{vazdekis:2015}
{Vazdekis} A.,  {Coelho} P.,  {Cassisi} S.,  {Ricciardelli} E.,
  {Falc{\'o}n-Barroso} J.,  {S{\'a}nchez-Bl{\'a}zquez} P.,  {Barbera} F.~L.,
  {Beasley} M.~A.,    {Pietrinferni} A.,  2015, \mnras, 449, 1177

\bibitem[\protect\citeauthoryear{{Vazdekis}, {S{\'a}nchez-Bl{\'a}zquez},
  {Falc{\'o}n-Barroso}, {Cenarro}, {Beasley}, {Cardiel}, {Gorgas} \&
  {Peletier}}{{Vazdekis} et~al.}{2010}]{vazdekis:2010}
{Vazdekis} A.,  {S{\'a}nchez-Bl{\'a}zquez} P.,  {Falc{\'o}n-Barroso} J.,
  {Cenarro} A.~J.,  {Beasley} M.~A.,  {Cardiel} N.,  {Gorgas} J.,    {Peletier}
  R.~F.,  2010, \mnras, 404, 1639

\bibitem[\protect\citeauthoryear{{Watkins}, {Evans} \& {An}}{{Watkins}
  et~al.}{2010}]{watkins:2010}
{Watkins} L.~L.,  {Evans} N.~W.,    {An} J.~H.,  2010, \mnras, 406, 264

\bibitem[\protect\citeauthoryear{{Wolf}, {Martinez}, {Bullock}, {Kaplinghat},
  {Geha}, {Mu{\~n}oz}, {Simon} \& {Avedo}}{{Wolf} et~al.}{2010}]{wolf:2010}
{Wolf} J.,  {Martinez} G.~D.,  {Bullock} J.~S.,  {Kaplinghat} M.,  {Geha} M.,
  {Mu{\~n}oz} R.~R.,  {Simon} J.~D.,    {Avedo} F.~F.,  2010, \mnras, 406, 1220

\bibitem[\protect\citeauthoryear{{Wyse}}{{Wyse}}{2010}]{wyse:2010}
{Wyse} R.~F.~G.,  2010, in {Cunha} K.,  {Spite} M.,   {Barbuy} B.,  eds,
  Chemical Abundances in the Universe: Connecting First Stars to Planets
  Vol.~265 of IAU Symposium, {How galaxies form: Mass assembly from chemical
  abundances in the era of large surveys}.
pp 461--469

\end{thebibliography}

\appendix
\section{{\rel} as a function of global {\sig}}
Here we provide the {\rel} plots for the entire sample discussed in the paper, analyzed in bins of increasing $\sigma_{local}$ (shown in the main text), $\sigma_{global}$, as well as the second velocity moment  $V_{rms}$, with the view to providing reference for future studies and comparisons involving a variety of object and data types. We note that the plots involving the three quantities are in excellent agreement.

\begin{figure*}
\centering
\includegraphics[width=1.35\columnwidth,angle=90]{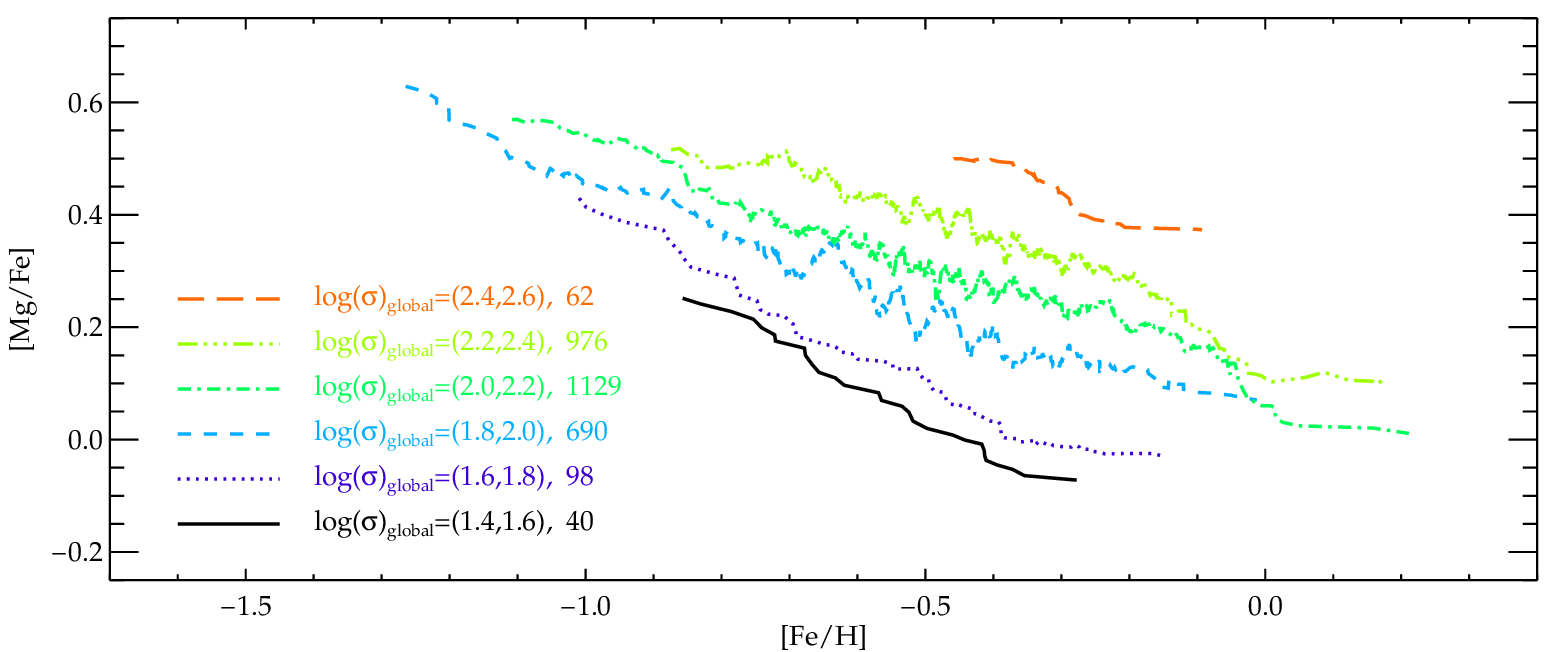}
\hspace{0.5cm}
\includegraphics[width=1.35\columnwidth,angle=90]{mgfe_vs_feh_averages_sauron_localsigma.eps}
\hspace{0.5cm}
\includegraphics[width=1.35\columnwidth,angle=90]{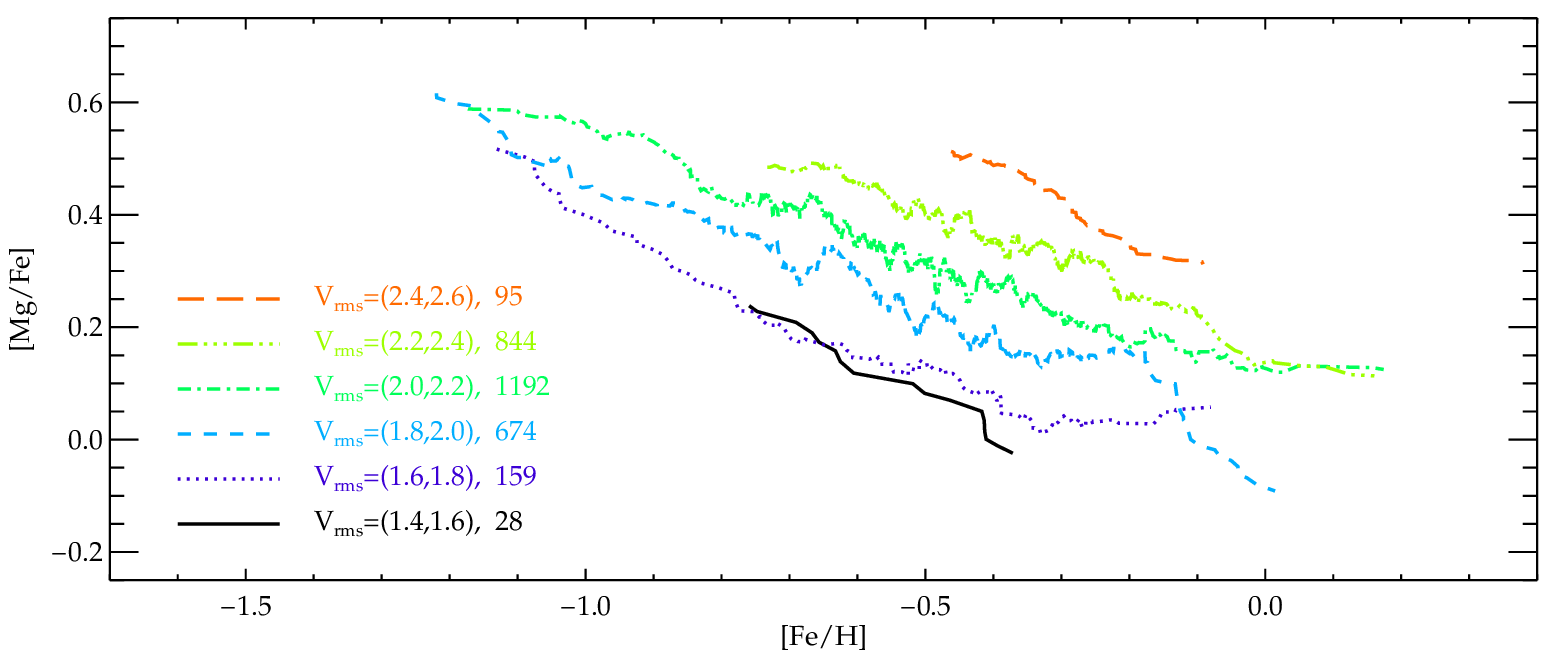}
\includegraphics[width=1.35\columnwidth,angle=90]{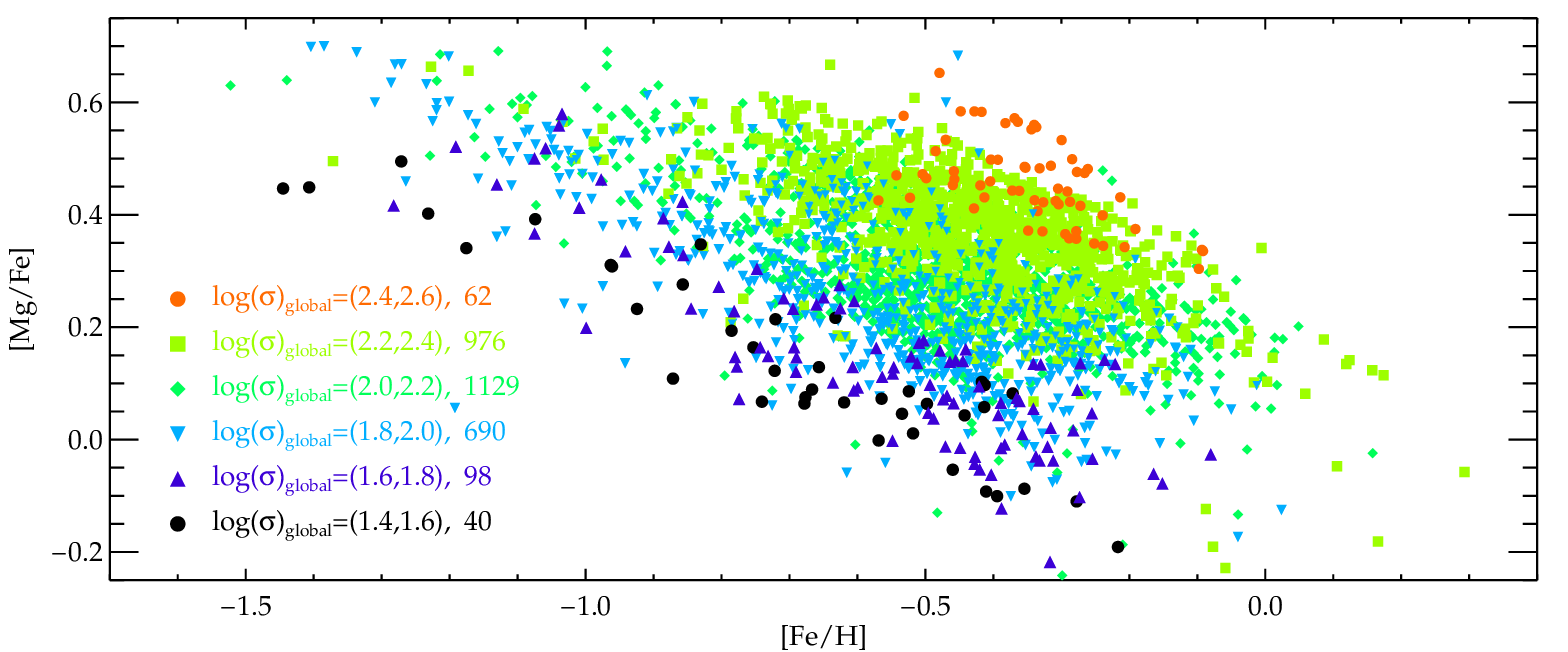}
\hspace{0.5cm}
\includegraphics[width=1.35\columnwidth,angle=90]{mgfe_vs_feh_allpoints_sauron_localsigma.eps}
\hspace{0.5cm}
\includegraphics[width=1.35\columnwidth,angle=90]{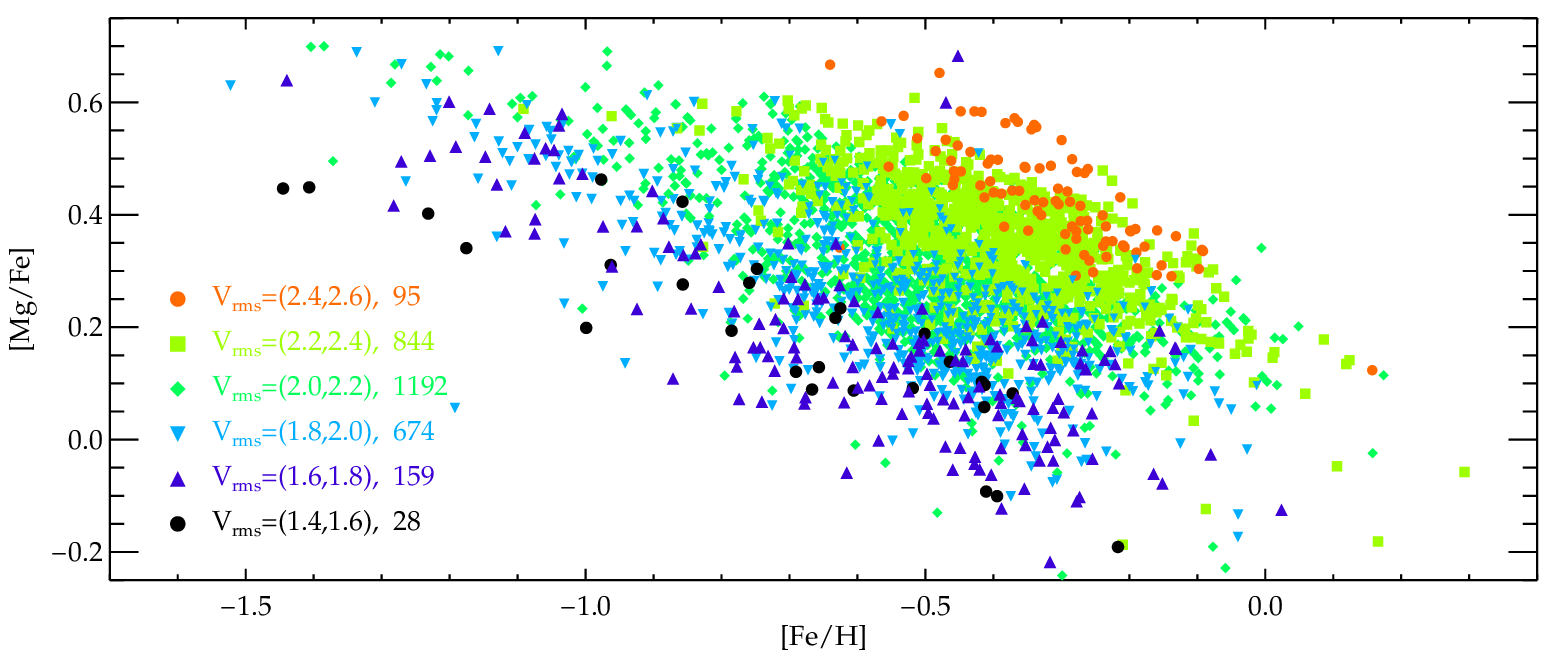}

\caption{As in Fig.\protect\ref{mgfe_vs_feh2} but for comparison color-coded by $\sigma_{global}$ (left), $\sigma_{local}$ (middle) and $V_{rms}$ values, For each of the quantities all individual profile points as well as running averages are shown.}
\label{mgfe_vs_feh3}  
\end{figure*}

\label{lastpage}

\end{document}